\documentstyle[seceq,epsf]{ptptex}

\newcommand\beq{\begin{equation}}
\newcommand\eeq{\end{equation}}
\newcommand\bea{\begin{eqnarray}}
\newcommand\eea{\end{eqnarray}}
\newcommand\tr{\mbox{tr\,}}

\title{
Classical fluctuations and semiclassical matrix elements
}

\author{
Bruno {\sc Eckhardt}$^1$,
Imre {\sc Varga}$^{1,2}$ and
P{\'e}ter {\sc Pollner}$^{1,3}$
}

\inst{
$^1$Fachbereich Physik, Philipps Universit{\"a}t Marburg,\\ 
D--35032 Marburg, Germany\\
$^2$Elm{\'e}leti Fizika Tansz{\'e}k, Fizikai Int{\'e}zet, Budapesti M\H{u}szaki
\'es Gazdas\'agtudom\'anyi Egyetem, H--1521 Budapest, Hungary\\
$^3$Komplex Rendszerek Fizik\'aja Tansz{\'e}k, E{\"o}tv{\"o}s Lor\'and 
Tudom\'anyegyetem, \\ H--1518 Budapest, Hungary
}

\recdate{
}

\abst{
We discuss the fluctuation properties of diagonal matrix elements 
in the semiclassical limit in chaotic systems. For extended observables,
covering a phase space area of many times Planck's constant,
both classical and quantal distributions are Gaussian. 
If the observable is a projection onto a single state or an
incoherent projection onto several states classical and quantal
distribution differ, but the mean and the variance are still
obtainable from classical considerations.}

\begin{document}

\maketitle

\section{Introduction}
\label{sec.1}

A central tool in the semiclassical analysis of chaotic systems is the
Gutzwiller trace formula which relates quantal properties to classical
periodic orbits \cite{Gutbook,Stoeckbook}. 
The extensions of Gutzwiller's trace formula that include matrix elements
in the density of states\cite{Wilkinson1,Eckhardt} allow to 
relate distributions of quantum matrix elements and distributions of 
classical trajectory segments \cite{Eckhardt95,EM}. Specifically,
quantum fluctuations are related to fluctuations in ensembles of 
classical trajectory segments of length Heisenberg time. As the 
classical limit is approached, the Heisenberg time diverges, and
the fluctuations vanish. This approach can therefore be used
\cite{Eckhardt95} to study the approach to ergodicity in quantum wave 
functions in chaotic systems. Integrable systems are not
ergodic on the energy shell and therefore the fluctuations do 
not vanish in the semiclassical limit, as shown by Mehlig \cite{Mehlig98}.
The form factors for the matrix element weighted 
densities of states have an additional constant contribution 
\cite{EM} and different behaviour in their rigidity \cite{physicad}.
Moreover, these relations can be used to analyse the effects 
near bifurcations \cite{VPE} and to study localisation effects
in wave functions. Beyond the static properties these can also be used 
to analyse the behaviour of cross sections~\cite{EFV,Dresden}.

The aim here is to analyse further an observation first made in the
calculation of cross sections. By Fermi's golden rule photodissociation
cross sections contain
matrix elements for observables that are projections onto 
initial states. In this limit
semiclassical arguments become suspect since the smoothness of
the observable that was assumed in the derivation of the trace formulas
is no longer guaranteed. However, numerical experiments indicate 
that even in this limit at least the widths of the distributions
can be obtained from classical considerations. This behaviour
of matrix elements of projection operators will be 
investigated further.

The behaviour of matrix elements was studied in a variety
of systems, including hydrogen in magnetic fields \cite{Eckhardt95,EM,Boose},
the bakers map\cite{Eckhardt95}, a perturbed cat map\cite{Carvalho}, 
and various billiards\cite{Mehlig98,Baecker}.
We here use the kicked standard map as the main example\cite{Izr90}. 
The main advantage of maps is that they allow
to study the semiclassical regime more easily and more deeply
than continuous systems.
 
The outline of the paper is as follows. In section \ref{sec.2} we summarise
information on the quantised standard map, on the semiclassical arguments
that relate classical and quantal matrix elements and on the 
behaviour expected from random matrix theory. The numerical
and analytical results on projectors are presented in section \ref{sec.3}.
We conclude with a summary and an outlook in section \ref{sec.4}.
 
\section{Preliminaries} 
\label{sec.2}

\subsection{The matrix element weighted density of states} 
In \cite{Eckhardt95} we derived formulae for the width of the  
quantum matrix element distribution in smooth systems and 
proposed a formula for maps by analogy to the smooth results. 
To set the stage for the present investigation and to explain 
the origin of this formula, we here sketch the equivalent 
derivation for quantised maps on a finite phase space
(similar to Carvalho et al\cite{Carvalho}).
 
Let $U$ be the unitary operator for the time evolution, $A$ 
the observable and $N_D$ the dimension of the Hilbert space. 
Then there are $N_D$ eigenphases $\phi_\mu$ and 
eigenstates $|\mu\rangle$, so that  
\beq 
U=\sum_{\mu=1}^{N_D} |\mu\rangle e^{i\phi_\mu} \langle \mu | \, . 
\eeq 
The matrix elements $A_\mu=\langle \mu|A|\mu\rangle$ are conveniently  
contained in the weighted density 
\beq 
\tilde\rho_A(\phi) = \sum_{\mu=1}^{N_D} A_\mu \delta(\phi-\phi_\mu) \,. 
\eeq 
It will be useful to work not with pure 
delta functions but with smeared out objects so that products  
are well defined. Rather than the Lorentzians or Gaussians used 
in previous works, the relation to finite trigonometric sums suggests
to work with sinc-functions, 
i.e. 
\beq
\delta_\epsilon(x) = {1\over 2\pi} {\sin[(2N+1)x/2] \over \sin(x/2)}
= {1\over 2\pi} \sum_{n=-N}^{N} e^{inx} 
\eeq
of width $\epsilon\approx2\pi/(2N+1)$. The advantage of this function 
is that the smoothed density can be expressed as a finite sum of 
traces of powers of the propagator, viz. 
\beq 
\rho_A(\phi) =  {1\over 2\pi} \sum_{n=-N}^{N}  
\sum_{\mu=1}^{N_D} A_\mu e^{in(\phi_\mu-\phi)} 
= {1\over 2\pi} \sum_{n=-N}^{N} \tr (A\,U^n) \, e^{-in\phi} \,. 
\eeq 
If $N$ is significantly smaller than $N_D$ some of the sinc-functions
will overlap and not all individual eigenphases 
can be identified. Very large 
$N$ are impracticable from the point of view of a semiclassical approximation 
to be made below. Thus a natural choice is $N$ of the order of $N_D$ so 
that all eigenstates can be resolved in the mean, 
yet the semiclassical approximations are still reasonable. 
 
The autocorrelation function of the smoothed density of states becomes 
\bea 
C_A(\theta) &=& {1\over 2\pi} \int_0^{2\pi}d\phi\,
\rho_A(\phi-\theta/2) \rho_A(\phi+\theta/2)\cr 
&=& {1\over 4\pi^2} \sum_{n=-N}^{N} \left|\tr (A\, U^n)\right|^2 e^{in\theta} 
\eea 
Up to some factors, the Fourier transform of this correlation function  
is the form factor. Here, since the spectrum is confined to the unit 
circle and a periodically continued interval in $\phi$, the  
form factor depends on a discrete argument only. We split off  
the density of states and define the form factor $K_n^{(A)}$ by 
\beq 
\tilde C_A(\theta) = {N_D\over 2\pi} {1\over 2\pi}  
\sum_{n=-\infty}^{\infty} K_n^{(A)} e^{in\theta} \,. 
\eeq 
When calculated with the smoothed density of states only the first 
few elements up to $N_D$ are included, but they do not differ from the  
exact ones, 
\beq 
K_n^{(A)} = {1\over N_D} \left|\tr (A\, U^n)\right|^2 \,. 
\eeq 
As in \cite{Eckhardt95}, all the information needed on 
matrix elements is contained 
in the semiclassical evaluation of $K_n$. 
 
Within the semiclassical approximation the trace is calculated from 
a stationary phase approximation to the propagator, which leads to  
an expression involving classical periodic orbits. Specifically, 
\beq 
\tr (A\, U^n) |_{sc} = 
\sum_{\gamma\in (n)} A_\gamma w_\gamma e^{iS_\gamma/\hbar} 
\eeq 
where the sum extends over all distinct orbits $\gamma$ which close  
after $n$ iterations. $S_\gamma$ is the action associated with the  
trajectory, $w_\gamma$ the amplitude including phases and  
$A_\gamma$ the sum of the observable $A$ over all points of the  
trajectory. Of the amplitudes we need that their modulus squared 
equals the classical periodic orbit weight, 
$|w_\gamma|^2 = 1/|\det (1-M_\gamma)|$,
where $M_\gamma$ is the linearisation perpendicular to the orbit. 
Thus the semiclassical form factor becomes 
\beq	
|\tr (A\, U^n)|_{sc}|^2 =  
\sum_{\gamma,\gamma'\in (n)} A_\gamma A^*_{\gamma'}  
w_\gamma w^*_{\gamma'} e^{i(S_\gamma-S_{\gamma'}/\hbar} 
\approx g \sum_{\gamma\in (n)} |A_\gamma|^2 |w_\gamma|^2 \,. 
\label{trausc}
\eeq 
The neglect of the contributions from orbit pairs $(\gamma,\gamma')$
with $\gamma\ne\gamma'$ constitutes the diagonal 
approximation\cite{Berry85}. The factor 
$g$ arises from the possibility of a quantum interference of classically 
distinct orbits with the same action, e.g. in the case 
of time reversal invariant systems, where a path and its time reversed 
image correspond to different objects in phase space, but with 
the same action. Because of the quantum interference between 
amplitudes, it contributes with twice the classical weight, hence 
$g=2$. In systems without such symmetries, like the unitary ensemble, $g=1$. 
 
The extraction of the variance of the matrix elements from the
form factors proceeds as in Ref.~\cite{EM}. The contributions 
$A_\gamma$ from orbits with similar period consist of two parts, 
one proportional to the mean of the observable and
to the length of the orbit and one capturing the fluctuations, 
i.e. $\langle A_\gamma^2\rangle = \langle A_\gamma\rangle^2 n^2
+ \alpha n$. Taking the square, averaging over orbits of similar
length and exploiting the periodic orbit sum rule (which absorbs one
factor of $n$) gives
\beq
K_n^{(A)} \approx g \langle A\rangle^2 \frac{n}{N_D}
+ g\alpha \frac{1}{N_D}
\label{kat}
\eeq
The linear increase of the first term is the small $n$ expansion
of the form factor, correct up to $n=N_D$ for the unitary
ensemble but approximate only for the orthogonal ensemble\cite{Haake}.
If the mean of the observable vanishes $K_n^{(A)}$ is already
the variance of the matrix elements and thus contained in the
second term of (\ref{kat}). Since the periodic orbit sum rule is an
asymptotic statement, we take $n$ as large as possible, but limited
by the requirement that the diagonal approximation be reasonable and
a semiclassical approximation possible. We therefore take $n=N_D$,
the Heisenberg time.

This calculation relates only the first and second moments of 
classical and quantal distributions. The main aim here will be
to compare the full distributions, especially in the limit of 
singular observables.

\subsection{Connection to random matrix theory} 

As demonstrated in a number of studies\cite{Izr90,GuhrPhy,BohigasLeH}, 
the short range statistical 
properties of quantised chaotic maps are in good agreement with 
those of ensembles of random matrices. So in the present context 
one can ask for the prediction of random matrix theory for the  
diagonal matrix elements. In particular, for the kind of maps 
and observables studied here, the relevant input is the statistics 
of eigenvector components. For the unitary ensembles, the distribution 
of real and imaginary parts can be calculated from the assumption 
of independent elements and the normalisation condition. For sufficiently 
large dimensions, this results in essentially Gaussian distributions, 
\beq 
P(x,y) = \sqrt{N_D\over\pi} e^{-N_D x^2}\,\sqrt{N_D\over\pi} e^{-N_D y^2} \,. 
\eeq 
If $A_{\mu}=\langle \mu|A|\mu\rangle$ denotes the matrix element of  
$A$ with respect to the basis set, the matrix element with 
respect to an eigenstate $\mu$ described by the amplitudes 
$x_\mu+iy_\mu$ becomes 
\beq 
A = \sum_\mu A_\mu (x_\mu ^2+y_\mu ^2)\, . 
\eeq 
The probability density for the distribution of the matrix elements  
then becomes 
\beq 
P(A) = \int \delta(A- \sum_\mu A_\mu (x_\mu ^2+y_\mu ^2))\,  
\prod_{\mu =1}^{N_D} P(x_\mu,y_\mu ) dx_\mu dy_\mu \,. 
\eeq 
By the usual manipulations this becomes a Gaussian in the limit
of large $N_D$,
\beq 
P(A) = \sqrt{N_D\over 2\pi \sigma_A^2} 
e^{-N_D (A-\langle A\rangle)^2/2 \sigma_A^2} 
\label{cldf}
\eeq 
where 
\beq
\langle A\rangle = {1\over N_D} \sum_\mu A_\mu \qquad
\sigma_A^2 = {1\over N_D} \sum_\mu A_\mu ^2 -\langle A\rangle^2
\label{cldf2}
\eeq
are the average and variance of the matrix elements 
of $A$ in the original basis. 

\subsection{The map}

We close this section with a description of the quantum version
of the kicked rotator used here.
The kicking potential
\beq 
V(\phi)=k(\cos\phi - \gamma\sin(2\phi)) 
\label{vfi}
\eeq 
gives rise to the classical map
\bea 
p_{n+1} &=& p_n + k(\sin\phi + 2\gamma\cos(2\phi)) \nonumber\cr 
\phi_{n+1} &=& \phi_n + p_{n+1} \, . 
\label{stmap}
\eea 
The parameter $k$ is the kick strength and $\gamma$ is 
used to break a reflection symmetry of the map. It is obviously 
periodic of period $2\pi$ in $\phi$. As we will study the quantum 
map at the case of resonance, the momentum $p$ is not unbounded 
but also restricted to be periodic with period $2\pi r$, where 
$r$ is the number of classical resonances. The parameter
$\gamma$ can be used to change continuously between the
cases with time reversal symmetry ($\gamma=0$) and without
(we take $\gamma=1$, for intermediate behaviour see \cite{BS,Carvalho}).

In the earlier quantum models of this map, phase space was taken to 
be an infinitely extended cylinder where arbitrarily large momenta 
were allowed,  the quantum map was infinite. However, it was noted 
that for rational values of $\hbar$, a consistent quantisation 
with a finite map can be achieved. If the classical phase space 
extends from $-\pi r$ to $+\pi r$ in momentum and if the quantum 
map contains $N_D$ states, the effective value of Planck's constant 
has a width of $2\pi$ in angle, the total phase space area 
is $4\pi^2 r$. This has to be divided into $N_D$ quantum cells, 
hence $h=4\pi^2 r/N$ or $\hbar=2\pi r/N_D$. We work in a momentum 
basis with wave functions 
$\langle \phi|n\rangle = e^{in\phi}/ \sqrt{2\pi}$
and momentum eigenvalues $p_n=\hbar n = 2\pi r n/N_D$. 
To avoid common factors between $r$ and $N_D$, we take 
the dimension odd, $N_D=2N_1+1$. The quantum unitary operator  
then has the form 
\beq 
U_{n,m} = {1\over N} e^{-{i\hbar\over 4}(n^2+m^2)} 
\sum_{j=-N_1}^{N_1} e^{-iV(\phi_j)/\hbar - i (n-m)\phi_j}
\label{evop} 
\eeq 
where the sum extends over the points $\phi_j=2\pi j/N_D$. 
In order to have full symmetry over the finite momentum interval, 
\beq 
U_{n+N_D, m} = U_{n,m+N_D} = U_{n,m}\,, 
\eeq 
one should take the number of resonances to be a multiple of $4$ 
(if $r$ is just even translation by $N_D$ produces a sign change). 
  
\section{Quantum and classical matrix element distributions} 
\label{sec.3}

In this section we compare the distributions of matrix elements 
in several cases. We consider the quantum kicked rotator 
(\ref{vfi}, \ref{evop}) and 
its classical counterpart (\ref{stmap}). The classical kicking 
strength is taken to be $K=k\hbar=7$ in order 
to ensure complete ergodicity over the whole phase space\cite{shepel}.
The matrix dimension $N_D=201$ turns out to be sufficient to provide
appropriate statistics.
 
\subsection{Chaotic delocalised states} 
\label{ergod}

The observable we consider is a projection onto a subset, $I(m)$, of the
unperturbed momentum basis states of length $m$,
\beq
A=\sum_{n\in I(m)}|n\rangle\langle n|.
\eeq
Its matrix elements in the basis of the eigenstates, $\mu$, of the 
evolution operator (\ref{evop}) are
\beq
A_\mu=\langle\mu|A|\mu\rangle =\sum_{n\in I(m)}|\langle \mu|n\rangle |^2.
\eeq
The projection covers a fraction $p=m/N_D$ of the phase space both
quantum mechanically and classically.

Due to normalisation of the eigenstates the mean value of $A_\mu$ is 
$\langle A\rangle =p=m/N_D$. The second moment, $\langle{A^2}\rangle$, is defined as
\beq
\langle{A^2}\rangle =\frac{1}{N_D}\sum_\mu\,\sum_{k,l\in I(m)}
     |\langle \mu|k\rangle |^2|\langle \mu|l\rangle |^2.
\eeq
which for $m\ll N_D$ can be calculated using the assumptions of random
matrix theory to give
\beq
\langle{A^2}\rangle =p^2 + g\frac{1}{N}p\, ,
\label{a2}
\eeq
where $g=1$ ($g=2$) for the case of unitary (orthogonal) symmetry.
Hence the variance is simply $\sigma^2_A=gp/N$.

If $m$ is sufficiently broad, $m\gg 1$, we may easily compare the distribution
of these quantum matrix elements to that of the classical distribution
which, in the limit $N>m\gg 1$ is essentially a Gaussian 
(\ref{cldf}, \ref{cldf2}). For the classical observable the mean and the 
fluctuating part can be calculated using a random--walk analogue due to the 
rapid decay of correlations between steps,
\beq
\langle A \rangle= p = m/N_D, \qquad
\alpha = p(1-p). 
\label{cl12}
\eeq
The classical fluctuations in the limit of $p\ll 1$ are related to the 
quantum variance~\cite{Eckhardt95}
\beq
\sigma_A^2=g\frac{1}{N}\alpha\,.
\label{cl-qm}
\eeq
\begin{figure}
   \epsfxsize = 10cm   
   \centerline{\epsfbox{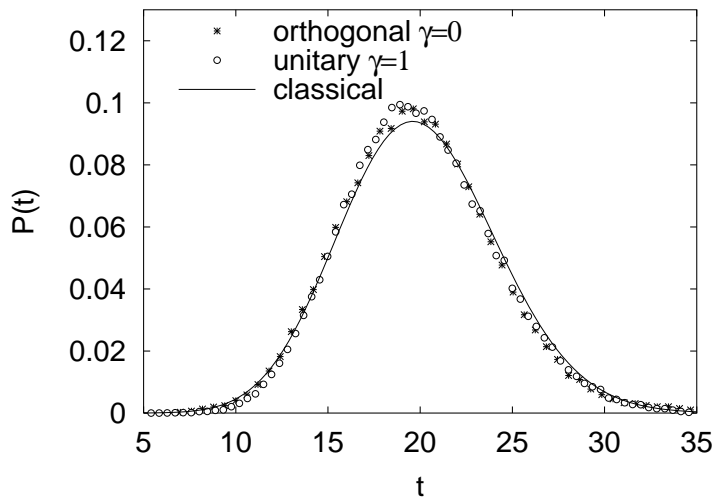}}
   \caption{Distribution of the matrix elements $A_{\mu}$ with $t=N_DA_{\mu}$.
The orthogonal symmetry ($\gamma=0$) has been rescaled according to 
(\protect\ref{gfac}). The continuous curve is the binomial distribution
as given in (\protect\ref{binom}). The parameters are: $N_D=201$ and $m=20$.}
   \label{clqu-distr}
\end{figure}
Fig.~\ref{clqu-distr} clearly demonstrates that the classical and the
quantum distributions are indeed very similar when the quantum kicked 
rotator is used in the unitary symmetry ($\gamma=1$). In the figure we also
plotted the distribution of matrix elements for $\gamma=0$, i.e. for the
orthogonal case. In the latter case we adopted a rescaling according to the
symmetry factor (\ref{cl-qm}) 
\beq
\sigma^2_A(\gamma=0)=g\,\sigma^2_A(\gamma=1).
\label{gfac}
\eeq
where the $g$--factor ($g=2$), as in (\ref{trausc}, \ref{kat}, \ref{a2},
\ref{cl-qm}) 
accounts for the fact that in the orthogonal 
symmetry the orbitals and their time reversed counterparts are 
indistinguishable. Results on the transition between the two ensembles
can be found in Bl\"umel and Smilansky \cite{BS} and in Carvalho et al.
\cite{Carvalho}

\begin{figure}
   \epsfxsize = 10cm   
   \centerline{\epsfbox{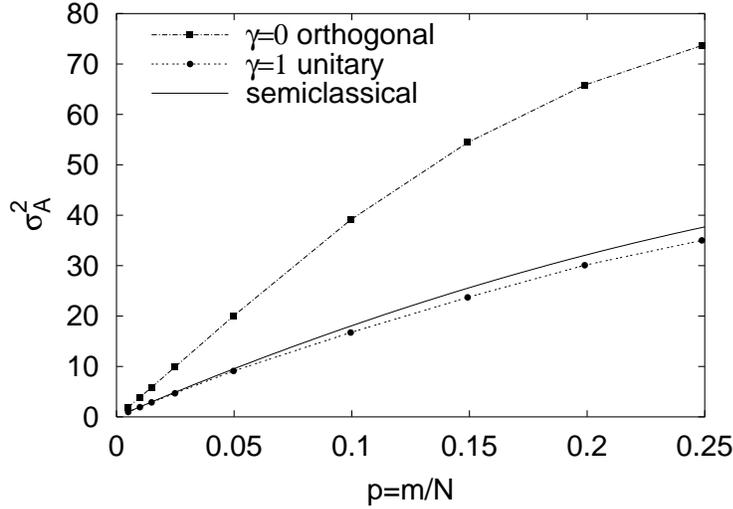}}
   \caption{Quantum versus semiclassical variance of $A_{\mu}$ as a 
function of its mean value $\langle A\rangle =m/N_D$. The classical curve is 
$\alpha N$ (\protect\ref{cl12}). We have used $N_D=201$.}
   \label{clqu-var}
\end{figure}

In fact one can look at the variance of the observable and compare it
to the mean value for different values of $m/N$ and compare with the classical
predictions. In Fig.~\ref{clqu-var} it is demonstrated that indeed for a wide
range of $m$ for a fixed value of $N_D=201$ the behaviour of the variance is
the same as presented also in Fig.~\ref{clqu-distr}.

\subsection{Singular observables} 
 
As the width of the projector, $m$, becomes smaller the assumption of  
an observable that is smooth on the scale of Planck's constant is
not satisfied anymore. Nevertheless, the data of Fig.~\ref{clqu-var} 
show that the variances are still in agreement with the semiclassical
predictions. The full distribution, however, cannot remain Gaussian 
for $m$ approaching one, neither classical nor quantal.
On the quantum side, the distribution becomes exponential for
the unitary ensemble and a Porter Thomas distribution 
$P(t)={e^{-t/2}}/{\sqrt{2t}}$ for the orthogonal ensemble. The classical 
distribution, in the case of strong chaos, is a binomial distribution, 
i.e. the probability to hit the support of the observable $k$-times in 
$N_D$ trials is
\beq
W_k(N_D)={N_D\choose k}p^k(1-p)^{N_D-k},
\label{binom}
\eeq
where $p=m/N_D$. From (\ref{binom}) $\langle A\rangle=Np=m$ and 
$\sigma^2_A=N\alpha=Np(1-p)$. The limit of $m$ approaching $1$ and diverging 
$N_D$ is the Poisson limit of the binomial distribution: for $p\to 0$ while
$N_D\to\infty$ so that $pN=m$ is fixed the above distribution reduces to
\beq
P_k=\frac{m^k}{k!}e^{-m}\,,
\label{poiss}
\eeq
with a clearly non-exponential dependence on $k$.
\begin{figure}[tbh]
   \epsfxsize = 10cm   
   \centerline{\epsfbox{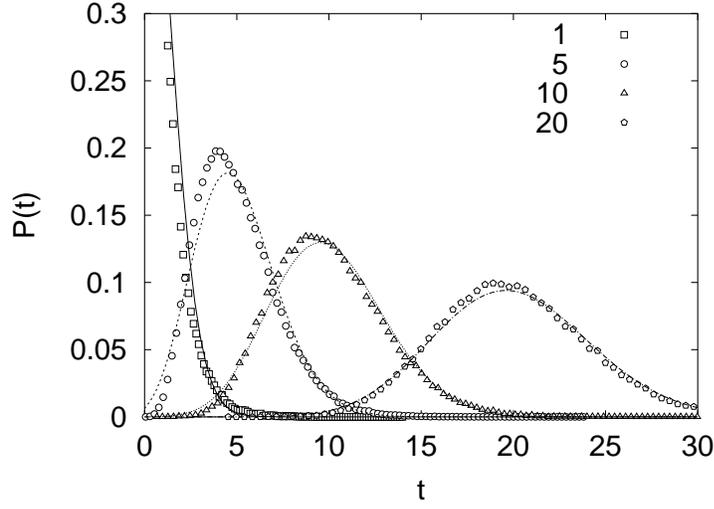}}
   \caption{Distribution of the matrix elements $A_{\mu}$ with $t=N_DA_{\mu}$
for the case of unitary symmetry ($\gamma=1$) with $m=1$, 5, 10, and 20.
The continuous curves are the classical binomial distributions 
(\protect\ref{binom}). We have used $N_D=201$.}
   \label{ch-m-1}
\end{figure}

\begin{figure}[tbh]
   \epsfxsize = 10cm   
   \centerline{\epsfbox{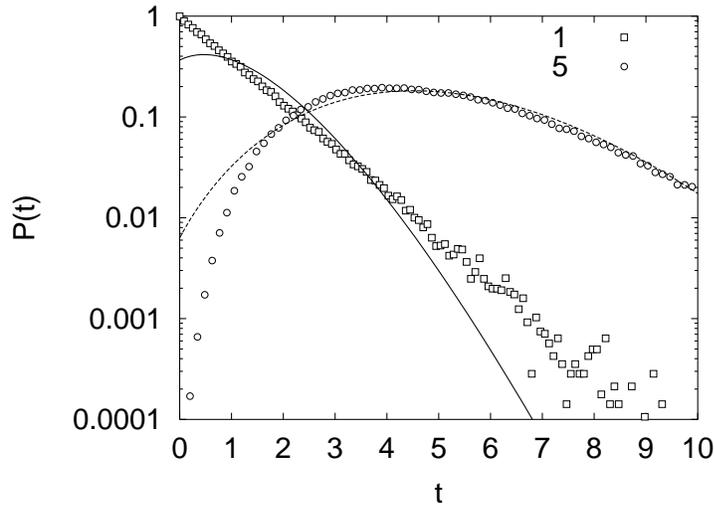}}
   \caption{Distribution of the matrix elements $A_{\mu}$ with $t=N_DA_{\mu}$
for the case of unitary symmetry ($\gamma=1$) with $m=1$, 5, compared to the
classical binomial distributions (\protect\ref{binom}). 
We have used $N_D=201$.}
   \label{low-m}
\end{figure}

\begin{figure}[tbh]
   \epsfxsize = 10cm   
   \centerline{\epsfbox{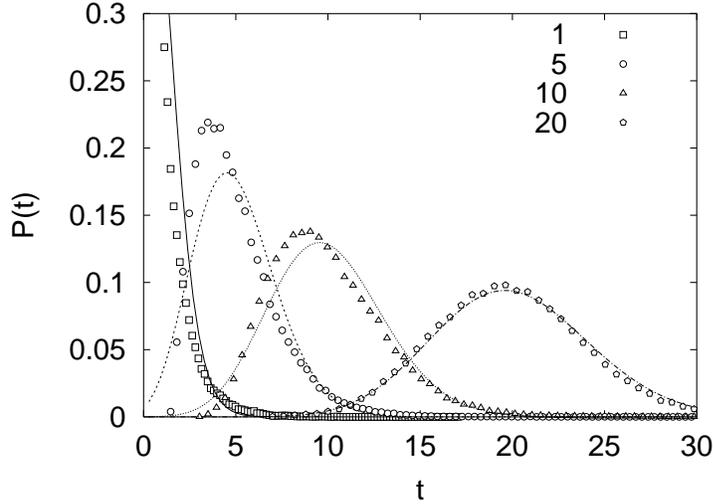}}
   \caption{Distribution of the matrix elements $A_{\mu}$ with $t=N_DA_{\mu}$
for the case of orthogonal symmetry ($\gamma=0$) with $m=1$, 5, 10, and 20.
The continuous curves are the classical binomial distributions
(\protect\ref{binom}). The numerical data have been rescaled according to
(\protect\ref{gfac}). We have used $N_D=201$.}
   \label{ch-m-0}
\end{figure}

Fig.~\ref{ch-m-1} shows the change of the distribution of matrix elements,
$A_{\mu}$, for different values of $m$ and fixed $N_D=201$. The 
histograms nicely follow the classical distribution for large $m$. To
highlight the differences for smaller values we show in Fig.~\ref{low-m}
the cases $m=1$ and $m=5$ on a semi-logarithmic plot. The exponential
distribution of the quantum matrix elements is clearly visible as is
the strongly non-Gaussian behaviour of the classical distribution.
The enhanced return probability in the orthogonal case and the singular
behaviour of the Porter Thomas distribution give result in more
pronounced differences between classical and quantal distributions
than in the unitary ensemble. For instance, the distributions for 
$m=5$ in Fig.~\ref{ch-m-0} for orthogonal symmetry
show larger differences than the corresponding ones in Fig.~\ref{ch-m-1}
for the unitary ensemble.

\subsection{Matrix element distributions near bifurcations} 

\begin{figure}[tbh]
   \epsfxsize = 10cm   
   \centerline{\epsfbox{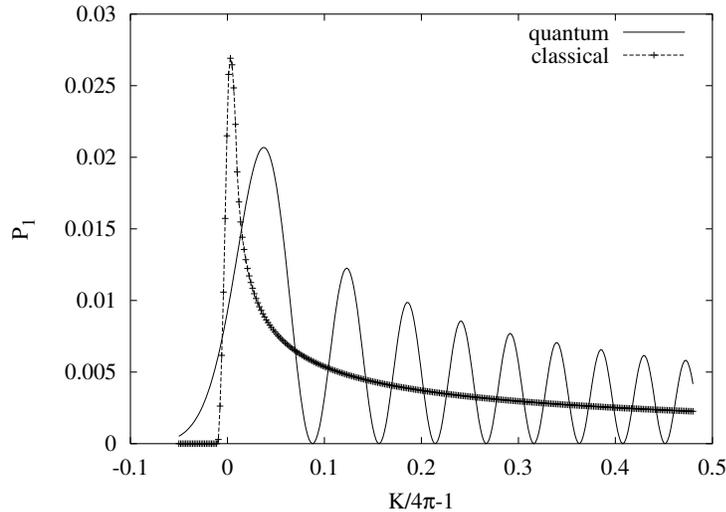}}
   \caption{Classical and quantum return probability to the neighbourhood
of bifurcation above $K=4\pi$ for $N_D=101$}
   \label{retprob}
\end{figure}

The previous sections have shown that while it is not always possible
to obtain the full distribution of matrix elements from the classical
distributions, at least the mean and the variance come out reliably.
This opens the possibility to study also the distributions in situations
where the assumptions of random matrix theory are not satisfied anymore,
specifically in systems with mixed phase space\cite{MME} or
near bifurcations of classical orbits\cite{VPE,PE}. 
To illustrate the kind of behaviour that has to be expected
near a bifurcation we consider an observable that is localised in 
a strip of width $2\pi/N$ around the the position of the bifurcation, $p=2\pi$. 
Near the bifurcation the probability to stay in this box is enhanced
and so will be the contribution of this state to the matrix elements.
It is thus useful to compare directly the classical and quantal probabilities
to return to this box. Therefore, we compare
$P_{cl}$, the classical probability for
trajectories that start in the strip and return after $N$ steps and 
$P_{qm}=\tr (A\, U^N)$ (Ref. \cite{PVE}). 
As can be seen from Fig.~\ref{retprob} both quantities show a qualitatively
similar behaviour, with a rapid increase near to
and a slower decline further above the bifurcation point. However,
the maxima are shifted in a characteristic fashion and the scaling
with $N$ is different\cite{PVE}.

\section{Conclusions} 
\label{sec.4}

The semiclassical theory used in Ref.~\cite{Eckhardt95} to relate 
matrix elements and classical trajectory segments requires the 
observable to be smooth on scales of Planck's constant. Nevertheless,
the examples discussed here show that this relation seems to hold
all the way down to singular observables like projection operators
onto single states. Quantum effects are always present through the 
symmetry factor $g$, but they are enhanced for singular observables
where the form of the distribution and thus in particular the higher
moments are different from what could be expected classically.
Nevertheless, the possibility to analyse the second moment
semiclassically should be helpful in analysing quantum chaos
in systems with mixed phase spaces or near bifurcations.

\section*{Acknowledgements}
We are grateful for financial support from the Alexander von
Humboldt Foundation, from DAAD-M\"OB within their 
Scientific Exchange Program and from the
Hungarian Committee for Technical Development (OMFB) through
the Grants Nos. OTKA T029813, T024136 and F024135.
We would like to thank M. Robnik for organising the stimulating
meeting where parts of these results were presented.



\end{document}